\documentclass[preprintnumbers, showkeys, article,floatfix,10pt,prd,onecolumn, superscriptaddress,nofootinbib,aps]{revtex4-2}
\usepackage{bm}
\usepackage{amsfonts}
\usepackage{latexsym}
\usepackage[latin1]{inputenc}
\usepackage{graphicx}
\usepackage{amsmath}
\usepackage{palatino}
\usepackage{mathpazo}
\usepackage{textcomp}
\usepackage{float}
\usepackage{booktabs}
\usepackage{dcolumn}
\usepackage{ragged2e}
\usepackage{hyperref}
\usepackage{amsmath}
\usepackage{xcolor}
\usepackage{orcidlink}
\usepackage{epsfig}
\usepackage[caption=false]{subfig}
\usepackage{commath}
\usepackage{multirow}
\usepackage{algorithm}
\usepackage{algpseudocode}
\usepackage{array}
\usepackage{silence}
\hypersetup{colorlinks,citecolor=blue}
\hypersetup{colorlinks=true,linkcolor=red,filecolor=magenta,    urlcolor=blue}

\newcommand{\be}{\begin{equation}}
\newcommand{\ee}{\end{equation}}
\newcommand{\bea}{\begin{eqnarray}}
\newcommand{\eea}{\end{eqnarray}}
\newcommand{\eeas}{\end{eqnarray*}}
\newcommand{\beas}{\begin{eqnarray*}}

\allowdisplaybreaks[1]

\begin{document}

\title{Simulation of a Variational Quantum Perceptron using Grover's Algorithm}
\author{Nouhaila Innan\orcidlink{0000-0002-1014-3457}}
\email{nouhailainnan@gmail.com}

\begin{abstract}
The quantum perceptron, the variational circuit, and the Grover algorithm have been proposed as promising components for quantum machine learning. This paper presents a new quantum perceptron that combines the quantum variational circuit and the Grover algorithm. However, this does not guarantee that this quantum variational perceptron with Grover's algorithm (QVPG) will have any advantage over its quantum variational (QVP) and classical counterparts. 
\newline Here, we examine the performance of QVP and QVP-G by computing their loss function and analyzing their accuracy on the classification task, then comparing these two quantum models to the classical perceptron (CP). The results show that our two quantum models are more efficient than CP, and our novel suggested model QVP-G outperforms the QVP, demonstrating that the Grover can be applied to the classification task and even makes the model more accurate, besides the unstructured search problems. 

\end{abstract}


\color{black} 
%


\affiliation{Quantum Physics and Magnetism Team, LPMC, Faculty of Sciences Ben M'sick,\\ Hassan II University of Casablanca, Morocco.}

\author{Mohamed Bennai\orcidlink{0000-0002-7364-5171}}
\email{mohamed.bennai@univh2m.ma}

\affiliation{Quantum Physics and Magnetism Team, LPMC, Faculty of Sciences Ben M'sick,\\ Hassan II University of Casablanca, Morocco.} 
\affiliation{Lab of High Energy Physics, Modeling, and Simulations, Faculty of Sciences,\\ University Mohammed V-Agdal, Rabat, Morocco.} %
\maketitle
\section{Introduction}
\indent Recently, there has been an increasing number of studies to combine the disciplines of quantum information and machine learning, and a variety of theories to merge these fields have consistently been put forward since machine learning is under pressure due to a lack of processing power of the increased amount of data in the world, and quantum computing offers these super computational capabilities. 
\\
\indent The combination of these two fields invariably leads to a massive interest in innovative information processing mechanisms that open up a new and improved range of solutions for various domains of applications, and the first concept was the research on quantum models of neural networks; it was essentially biologically inspired, in the hope of finding explanations for brain function within the framework of quantum theory \cite{A1}.
In 2013, this combination got the name quantum machine learning by Lloyd et al. \cite{A2} as a definition of an area of research that explores the combination of quantum information and ML principles.
\\
\indent However, the development of potential quantum machine learning algorithms has made some progress; several famous classical ML algorithms already have quantum analogs, such as the quantum support vector machine (QSVM), quantum k-means clustering, quantum Boltzmann machine (QBM), and the quantum perceptron (QP) which there have been some papers that mainly overview methods and algorithms of this model.
\\
\indent Zhou et al. \cite{A3} developed a quantum perceptron approach based on the quantum phase capable of computing the XOR function using only one neuron, then Siomau et al. \cite{A4} introduced an autonomous quantum perceptron based on calculating a set of positive valued operators and valued measurements (POVM), after that Sagheer and Zidane \cite{A5} proposed a quantum perceptron based on Siomau method capable of constructing its own set of activation operators to be applied widely in both quantum and classical applications to overcome the linearity limitation of the classical perceptron
\\
\indent In 2018, a multidimensional input quantum perceptron (MDIQP) was proposed by Yamamoto et al. \cite{A6}; their model had an arbitrary number of inputs with different synaptic weights, being able to form large quantum artificial neural networks (QANNs). And after that, Torrontegui and Ripoll suggested a unitary quantum perceptron as an efficient universal approximator using the sigmoid function with the possibility to apply it to different applications like quantum sensing \cite{A7}.
\\
\indent Wiebe et al. \cite{A8} introduced two quantum perceptron models based on Grover's search algorithm to minimize the error of QP, and based on that, we got inspired to think about Grover's algorithm as a way to develop our model. However, recently several studies showed that variational algorithms are so suitable for quantum machine learning models, especially the quantum perceptron \cite{A9, A10}, so in this work, we would like to provide a different way to implement this model by increasing its accuracy using the variational circuit and Grover's algorithm. The rest of this paper is organized as follows: in section \ref{section:my}, we review some basic knowledge of Grover's algorithm for unstructured search, and we describe the principal features of the classical perceptron by showing the mechanics of the algorithm; in Section \ref{section:my1}, we study the concept of a quantum variational perceptron by describing the state preparation, the model representation for the associated quantum circuit, and the measurement component. In Section \ref{section:my2}, we describe the quantum variational perceptron model with Grover's algorithm, and in Section \ref{section:my3}, we examine how well our model performs. Finally, in the last section \ref{section:my4}, this paper concludes with final remarks and future work.

\section{Background}
\label{section:my}
\subsection{Grover's Algorithm}

\begin{figure}[h]\centering
\includegraphics[scale=0.75]{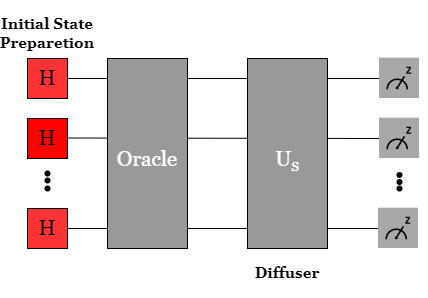}
\caption{An Overview of Grover's Algorithm Steps}
\label{fig1}
\end{figure}

\indent Grover's algorithm is a quantum algorithm for searching an unsorted database with $N$ items in a short amount of time. Usually, it would take $O(\sqrt{N})$ time since we would have to search through all the entries to find the right one \cite{A11}. Even though it is simply a quadratic speed-up, it is substantial when it is significant. Unlike many previous quantum algorithms, which address a "black box" problem, Grover's Search Algorithm solves a searching problem in which the objective is to achieve a particular state with a measurement among many possible states \cite{A12}. 
In the simplest form, the algorithm allows us to estimate (into the database) when we give a function (output of database), as shown the figure \ref{fig1}.
\\
\indent The steps of the Grover algorithm are as follows; first, we have to define input states. Without loss of generality, we will assume that the input states are integers between $0$ and $2^{n}$, where $n$ is an integer. The $2^{n}$ integer states will be encoded using the states $\vert0\rangle$ and $\vert1\rangle$ of $n$ qubits.
\\
\indent Second, for our case, we must define an Oracle function $f(x)$, which is a function that returns zero for all possible input states except one input state, and it should be encoded in an operator $O$ that acts as $O\vert x\rangle=\left ( -1 \right )^{f(x))}\vert x\rangle$ \cite{A13}, which means that the Oracle negates the probability amplitude of the input state $\vert x\rangle$ if and only if $f(x)=1$, and for better understanding, algorithm \ref{grovt} explains these steps:
\begin{algorithm}[H] $\bullet$\ \textbf{Step 1:} Initialization of the qubits in the $\vert0\rangle$ state and creation of a uniform superposition of all basis inputs.\\
\qquad$\bullet$\ \textbf{Step 2:} Execution of the Oracle.\\
\qquad$\bullet$\ \textbf{Step 3:} Application of Grover's diffusion operator (inversion about the mean).\\
\qquad$\bullet$\ \textbf{Step 4:} Repetitions of steps 2 and 3.\\
\qquad$\bullet$\ \textbf{Step 5:} Final measurement.
\caption{Grover's Algorithm}
\label{grovt}
\end{algorithm}

\subsection{Classical Perceptron} 
\begin{figure}[ht]\centering
	\includegraphics[scale=0.5]{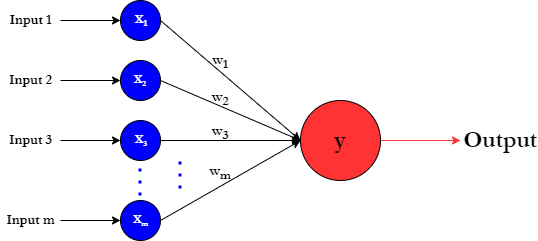}
	\caption{Graphical representation of perceptron} 
        \label{fig2}                 
\end{figure}

\indent Perceptron is an artificial neuron, and so it is a neural network unit. It performs calculations to detect features or patterns in the input data. It is an algorithm for supervised learning of binary classifiers. This algorithm allows artificial neurons to learn and process features in a data set \cite{A14}. Its modeling function is given by:
\begin{equation}
f(x;y)=\phi(x,x)
\end{equation}
\indent Where $x$ and $y$ are the inputs and outputs are real or binary numbers. Sometimes the mathematical structure makes it convenient to choose $\{ -1,1 \}$ rather than $\{ 0,1 \}$, and $\phi$ is the activation function referred to as the sign function or Heaviside function \cite{A15}. \\
\indent The perceptron plays an essential role in machine learning projects. It is frequently used to classify data or simplify or supervise binary classifier learning capabilities. Recall that supervised learning consists of teaching an algorithm to make predictions, which can be achieved by feeding the algorithm with already correctly labeled data. \\
\indent  And to better understand how it works, figure \ref{fig2} shows a perceptron in the typical neural network graph representation, where the inputs and outputs are considered units with certain values, updated by the units that influence them, and the connections between them are associated with a weight.

\subsubsection{Perceptron Algorithm}

\indent The conventional perceptron algorithm is designed for binary classification; consequently, we represent the training data used in this section as $S=\left \{ \left ( x_{i}, y_{i}  \right ) \right \}_{i=1}^{n}$ with $ x_{i} \in R $ and $ y_{i} \in \{ -1,1 \} $, the outputs $ y_{i} $ as mentioned before, can only accept two values \( 1 \, or - 1 \) hence the name binary classification \cite{A16}. \\
\indent A binary perceptron is a linear classifier. Given the data S, the objective is to learn the weight vector $ w \in R $ such that $ \left \langle w, x_{i} \right \rangle > 0 $ for any $ x $ belonging to class 1 and $ \left \langle w, x_{i} \right \rangle \leq  0 $ for any $ x $ belonging to class $ -1 $. \\
\indent The idea of the perceptron algorithm is to initialize $w$ arbitrarily, iterate several times (set a priory or until convergence) on the training data, and adjust the weight w each time a data element is misclassified, and we can formulate it as follows (Algorithm \ref{algi}):

\begin{algorithm}[H]
\caption{Perceptron Algorithm}
\begin{algorithmic}
\State Initialize the weights $w$ and the bias $b$ randomly. 
\For {to $n$}
    \For {each example $(x_{i}, y_{i})$ }
         \State Compute the prediction $y_{i}=sign\left( \left\langle w,x_{i} \right\rangle \right)$ 
         \If {$\hat{y_{i}} \neq y_{i}$} 
              \If {$y_{i}$ is positive}
                  \State Adjust $w: w=w+x_{i}$
              \ElsIf{$y_{i}$ is negative}
                  \State Adjust $w: w=wix_{i}$  
             \EndIf
          \EndIf
    \EndFor
\EndFor
\end{algorithmic}
\label{algi}
\end{algorithm}

\section{Quantum Variational Perceptron}
\label{section:my1}
\indent The Quantum Variational Perceptron (QVP) is a machine learning model that combines quantum computing and artificial neural networks. It is based on the variational quantum circuit (VQC) model \cite{A17, A18}, which shows how quantum circuits can represent the parameters of a neural network. The QVP uses quantum circuits to do complicated calculations that are hard to do classically, and the classical optimizer adjusts the parameters of these quantum circuits to make them optimized. 
\newline \indent One of the main advantages of the QVP is that it can handle high-dimensional and non-linear data, which is difficult to process with classical neural networks. The QVP can also do quantum-enhanced feature extraction to improve the model's accuracy, which is essential for natural language processing, drug discovery, and classification.  
\newline \indent The QVP also has potential advantages over classical neural networks in terms of robustness to noise and generalization that we will show in the results paragraph \ref{section:my3}, quantum computing devices are sensitive to noise, but the QVP can use quantum error correction techniques to mitigate this issue. This robustness to noise makes the QVP well-suited for applications where the data is uncertain or noisy.
\newline \indent So our goal is to use quantum computing to compute $f(x;\theta)$ and then use a classical optimizer to optimize the lost function, so we represent our problem with a quantum circuit \cite{A19, A20, A21}. We build this circuit in three steps: state preparation, model circuit, and measurement.

\subsection{State Preparation:}

\indent State preparation is an essential step in the quantum variational circuit \cite{A22}. It involves preparing the initial quantum state as an input to the rest of our circuit, and we can determine this state by a set of parameters that we vary to optimize the circuit's output \cite{A23}. There are many different approaches for state preparation, but one popular method is to use a series of unitary transformations to transform a known initial state into the desired variational state. These unitary transformations can be implemented using a variety of quantum gates, such as rotation gates, controlled-not gates, and phase gates. It is essential to carefully choose the initial state and the unitary transformations to maximize the chances of obtaining the desired output from the quantum circuit. 
\newline \indent So our first step is to encode the classical data $x$ into quantum data; as we mention, there are different approaches to accomplish it, like basis encoding, angle encoding, higher-order encoding, and amplitude encoding \cite{A24}.
\newline \indent Each methodology possesses a unique application. For instance, in the basis encoding method, the quantum state is represented on a specific basis, including the computational or Fourier bases. Conversely, in the angle encoding method, the quantum state is expressed in terms of the angles of rotation applied to the state. Additionally, in higher-order encoding, the quantum state is represented utilizing higher-order properties, such as entanglement or quantum coherence.
\newline \indent In our case, we chose the amplitude encoding method, which means that our data are directly associated with the amplitude of the quantum state; the idea is to represent these quantum states as complex-valued amplitudes to describe the probability of measuring a particular outcome when the quantum state is measured.
\newline \indent In this method, we represent quantum states as vectors within a complex vector space and quantum operations as matrices acting upon these vectors. This representation offers a compact and intuitive framework for describing quantum states and operations, making it particularly useful within the context of quantum variational circuits. The advantage of this method is the ability to encode a data set of $M$ inputs with $N$ features need only $n=log(N*M)$ qubits; it's based on creating an operator $\phi(x)$ which will result in the state $\phi(x) |0_p\rangle$ with $p$ the number of qubits. 
\subsection{Model Circuit:}
The model circuit step is a crucial component in the optimization process. In this step, a parametrized quantum circuit called an ansatz, in the form of a unitary operator, is built to model our quantum state. The parameters of the circuit are adjusted iteratively to minimize the difference between the model state and the target state. The model circuit step is essential for achieving accurate and efficient quantum state preparation and algorithm implementation. Also, the performance and success of our circuit depend on how well we choose the ansatz structure and the optimization algorithm.
\newline
\indent So we have built $U(\theta)$, a unitary operator from the quantum state $\vert\phi(x)\rangle$, which represents the vector $x$ in the quantum circuit, such that $U(\theta)\vert\phi(x)\rangle$ is a quantum state measurable, with $\theta$ the trainable parameter. \\
\indent After that, we decomposed $U(\theta)=U_{1}...U_{L}$ as a product of two qubits parameterized gates, then we used a combination of generic unitary gates and CNOT gates, and as a result, we have the expression of $U$ as follows:
\begin{equation}
  U(\theta, \phi, \lambda) = \begin{bmatrix}
  cos(\frac{\theta}{2}) & -e^{i\lambda}sin(\frac{\theta}{2})\\ 
  e^{i\phi}sin(\frac{\theta}{2}) & e^{i\lambda+i\phi}cos(\frac{\theta}{2})
\end{bmatrix} 
	\label{eq:refname2}
\end{equation}
\subsection{Measurement:} 
\begin{center}
\begin{figure}[htbp]\centering
	\includegraphics[width=\linewidth]{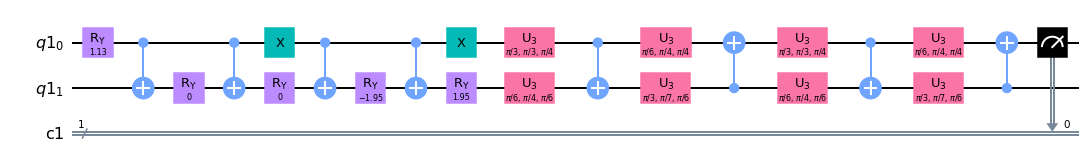}
	\caption{Quantum Variational Perceptron Circuit}
 \label{fig3}
\end{figure}
\end{center}

The measurement step allows the extraction of information from the quantum state, it involves the projection of the quantum state onto a measurement basis, and the collection of the resulting measurement outcomes is used to calculate the loss function; we use this function to adjust the parameters of our quantum circuit in this current step.  
\newline
\indent So we start by generating the first qubit of the quantum state using the circuit in figure \ref{fig3}, then we measured it, and it gave us this expression: 
\begin{align}
f(x;\theta) &= \mathbb{P}(y|x=1;\theta)= \mathbb{P}(q_0=1|x;\theta) = \sum_{k=1}^{n} |(U(\theta)\phi(x))_k|^2 
\label{eq3}
\end{align}
\indent We have simplified the expression \ref{eq3} using $\pi(x;\theta) $ as $ \mathbb{P}(q_0=1|x;\theta)$, now after we have set up all the parameters required, our model is trained to minimize a loss function which can be defined as follows:
\begin{equation}
  \mathcal{L}(\theta) = \frac{1}{N} \sum_{i=1}^{N} l(\pi(x_i;\theta), y_i)  
\end{equation}
\indent The parameters are updated via batch stochastic gradient descent; the difficulty was accurately calculating the gradient. Fortunately, it is possible to evaluate the gradient via quantum circuits, and as a result, we have equation \ref{eqgrad}:
\begin{equation}
    \nabla_\theta\mathcal{L}(\theta) = \frac{1}{N} \sum_{i=1}^{N} \nabla_\theta(l(\pi(x_i;\theta), y_i)) = \frac{1}{N} \sum_{i=1}^{N} \nabla_\theta\pi(x_i;\theta)\partial_1l(\pi(x_i;\theta), y_i)
 \label{eqgrad}
\end{equation}
Where $\partial_1l$ is the partial derivative of $l$ regarding the first variable, the expression of gradient loss became: 
\begin{equation}
    \nabla_\theta\pi(x_i;\theta) = \frac{-\nabla_\theta\mathbb{E}(\sigma_z)}{2} = -\frac{1}{2} \nabla_\theta\langle \phi(x)U | \sigma_z |U\phi(x) \rangle
\end{equation}
with $\sigma_z \otimes \mathbb{I} \otimes ... \otimes \mathbb{I}$ is abbreviated to $\sigma_z$.
After considering $\nu$ as an element of the vector $\theta$, we can represent the $\partial_\nu \pi(x;\theta)$ as: 
\begin{equation}
\partial_\nu\pi(x;\theta) = -\frac{1}{2} \langle \phi(x)\partial_\nu U| \sigma_z |U\phi(x) \rangle -\frac{1}{2} \langle \phi(x)U |\sigma_z |\partial_\nu U\phi(x) \rangle 
\label{epi}
\end{equation}
And using the rules of derivation, equation \ref{epi} became equal to the expression below (\ref{refn1}). 
\begin{equation}
\label{refn1}
\partial_\nu\pi(x;\theta) = -\frac{1}{2} (\langle \phi(x)\partial_\nu U| \sigma_z |U\phi(x) \rangle + \langle \phi(x)\partial_\nu U| \sigma_z |U\phi(x) \rangle^*) 
 = - Re\{ \langle \phi(x)\partial_\nu U| \sigma_z |U\phi(x) \rangle\}
\end{equation} 
\indent Since the $\partial_\nu U$ is not a unitary operator, we cannot use it in our quantum circuit; however, we achieve it due to our circuit because the derivative of $U$ only concerns the derivative of the elementary gate, which means that $\partial_\nu U = U_1...\partial_\nu(U_i)...U_L$ where $U_i$ is the gate to which $\nu$ belongs. For an elementary unitary gate, as described above, we have the following identities:
\begin{align}
\partial_\theta U = \frac{1}{2} U(\theta + \pi, \phi, \lambda)
\end{align}
\begin{align}
\partial_\phi U = \frac{i}{2} (U(\theta, \phi, \lambda) - U(\theta, \phi + \pi, \lambda))
\end{align}
\begin{align}
\partial_\lambda U = \frac{i}{2} U(\theta, \phi, \lambda) - U(\theta, \phi, \lambda + \pi))
\end{align}
\indent Therefore $\partial_\nu U$ can be computed as the form:
\begin{align}
 \sum_{k=1}^{K}a_k Re\{ \langle \phi(x)U(\theta^{[k]})| \sigma_z |U(\theta)\phi(x) \rangle\} + \sum_{l=1}^{L}b_l Im\{ \langle \phi(x)U(\theta^{[l]})| \sigma_z |U(\theta)\phi(x) \rangle\}
 \label{eq12}
\end{align}
where $\theta^{[k]}$ and $\theta^{[l]}$ are the modified vector of parameters from the above identities, the imaginary part comes from the $i$ in the same identities for $\phi$ and $\lambda$ with the result that we can calculate our loss function using \ref{eq12}.
\section{Quantum Variational Perceptron with Grover algorithm}
\label{section:my2}
According to the work of Khanal et al. \cite{A25}, in order to use Grover's algorithm in the classification task, they reformulated the problem of classification as a search problem, replacing Grover's oracle with the variational algorithm and adding quantum gates (AND, XOR, and OR). In order to introduce the notion of classification in Grover's algorithm, they used the Kernel method, which is a method that is more adapted to complex problems \cite{A26}; in addition, other research works try to achieve the classification task by applying the Grover algorithm in different methods \cite{A27, A28}. 
\newline
\newline

\begin{figure}[ht]\centering
\includegraphics[width=\linewidth]{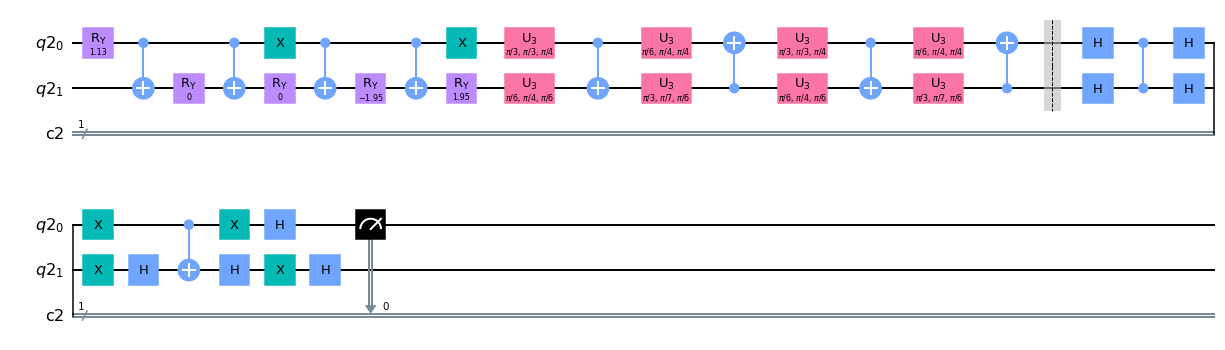}
\caption{Quantum Variational Perceptron-Grover Circuit}\label{fig4}
\end{figure}

\indent This paper suggests a new method based on adding the Grover circuit after the variational algorithm. As known, we can use the Grover circuit for amplitude amplification, so in our suggested method, we use it to speed up the training process by amplifying the amplitude of the target state. Thus making it easier to identify during the measurement step, besides amplifying the amplitude of the correct output state in the superposition of all possible output states and updating the weights to get a more accurate model. 
\\
\indent
The circuit is almost the same as the variational circuit. However, we obtained the classification by measuring the probability of the qubits in a particular configuration, and we executed the same circuit above several times to obtain the loss function values. 
We counted how many times the configuration appeared in the classified data set. It is two-dimensional because we use two qubits; we assign values to the parameters of the circuit, then we rescale these input values to fit the angles.
\\
\indent
Figure \ref{fig4} shows the quantum circuit applied in this method, using the same steps as in Section \ref{section:my2}, then adding a Grover circuit using Hadamard gates, CNOT gates, controlled Z gates, and Pauli X gates \cite{A29, A30}.  

\section{Results and Discussion}
\label{section:my3}
To test our model implementation, we loaded the Iris data set to evaluate its performance. We have calculated the accuracy for each model's iteration: the quantum variational perceptron with Grover and the variational perceptron using the \emph{ibmq\_qasm\_simulator}, which is one of the IBM quantum simulators \cite{A31}. 
This QASM simulator allows us to generate quantum circuits both ideally and subject to noise modeling with a maximum of $32$ qubits. 
The results verify if our suggested model fits the training and validation sets by using the result of the variational perceptron as a reference and comparing both quantum models to the classical model to show the quantum advantage. 
\newline
\indent Table \ref{tab1} shows how we summarized the results of the three different models, and comparing the accuracy of a model on the training set versus the validation set can give insight into the model's ability to generalize to new data. In general, we expect the model's accuracy on the training set to be higher than its accuracy on the validation set because the model has seen the training data during training and has learned to classify it correctly, and as expected, the table \ref{tab1} verified this.
\begin{table}[H]
\centering
\begin{tabular} {||c c c c c c c||}
\hline
 & \multicolumn{2}{|c|} {CP} &\multicolumn{2}{|c|} {QVP} &\multicolumn{2}{|c|} {QVP-Grover} \\
\hline
 Iter & Acc train & Acc valid & Acc train &  Acc valid & Acc train &  Acc valid \\
\hline 
 1 & 0.49 & 0.81  & 0.45 & 0.36  & 0.47 & 0.40 \\
 10 & 0.91 & 0.92  & 0.53 & 0.36  & 0.47 & 0.40  \\ 
 15 & 0.93 & 0.93  & 0.53 & 0.34  & 0.49 & 0.45  \\ 
 20 & 0.94 & 0.94  & 0.53 & 0.32  & 0.50 & 0.48  \\ 
 25 & 0.94 & 0.94  & 0.26 & 0.40  & 0.55 & 0.50  \\ 
 30 & 0.95 & 0.94  & 0.29 & 0.46  & 0.56 & 0.53  \\ 
 35 & 0.95 & 0.95  & 0.62 & 0.50  & 0.59 & 0.58  \\ 
 40 & 0.95 & 0.95  & 0.68 & 0.66  & 0.64 & 0.60 \\ 
 45 & 0.96 & 0.95  & 0.73 & 0.66  & 0.77 & 0.73  \\  
 50 & 0.96 & 0.95  & 0.84 & 0.66  & 0.85 & 0.80 \\ 
 55 & 0.96 & 0.96  & 0.99 & 0.66  & 0.91 & 0.90  \\ 
 60 & 0.96 & 0.96  & 0.94 & 0.60  & 0.97 & 0.91  \\ 
 65 & 0.96 & 0.96  & 0.96 & 0.70  & 1 & 0.91  \\ 
 70 & 0.96 & 0.96  & 0.96 & 0.74  & 1 & 0.94  \\ 
 75 & 0.97 & 0.96  & 0.98 & 0.81  & 1 & 0.97  \\  
 80 & 0.97 & 0.96  & 1 & 0.82  & 1 & 0.99  \\ 
 85 & 0.97 & 0.96  & 1 & 0.82  & 1 & 0.99  \\
 90 & 0.97 & 0.96  & 1 & 0.84  & 1 & 0.99  \\ 
 95 & 0.97 & 0.96  & 1 & 0.90  & 1 & 0.99  \\
 100 & 0.97 & 0.96  & 1 & 0.90  & 1 & 0.99 \\
\hline   
\end{tabular}
\caption{Summary of classification accuracy on the training and validation sets for Classical Perceptron (CP), Quantum Variational Perceptron (QVP) and Quantum Variational Perceptron with Grover(GVP-G).}\label{tab1}
\end{table}
\begin{figure}[!htb]
\minipage{0.5\textwidth}
  \includegraphics[width=\linewidth]{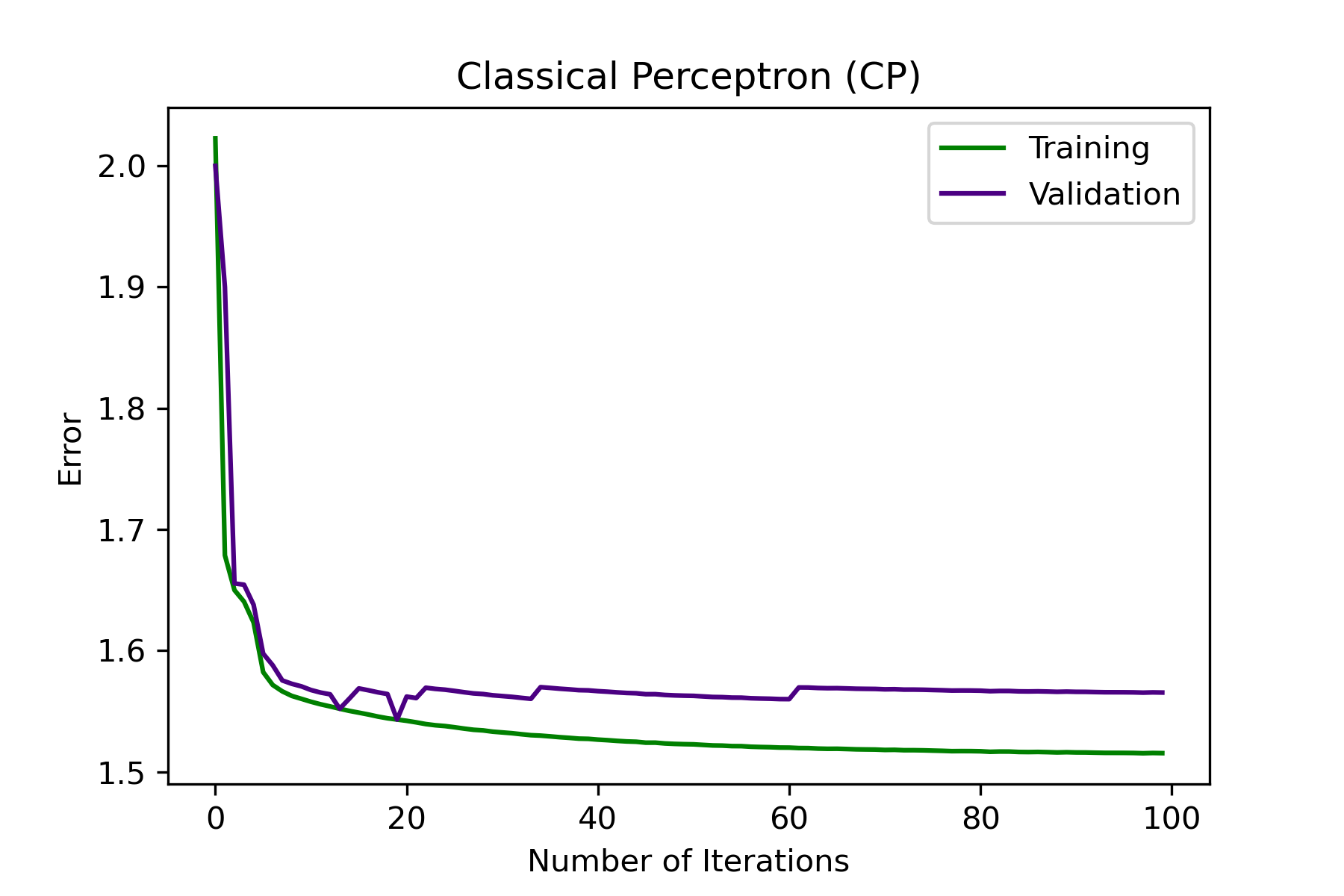}
  
\endminipage\hfill
\minipage{0.5\textwidth}
  \includegraphics[width=\linewidth]{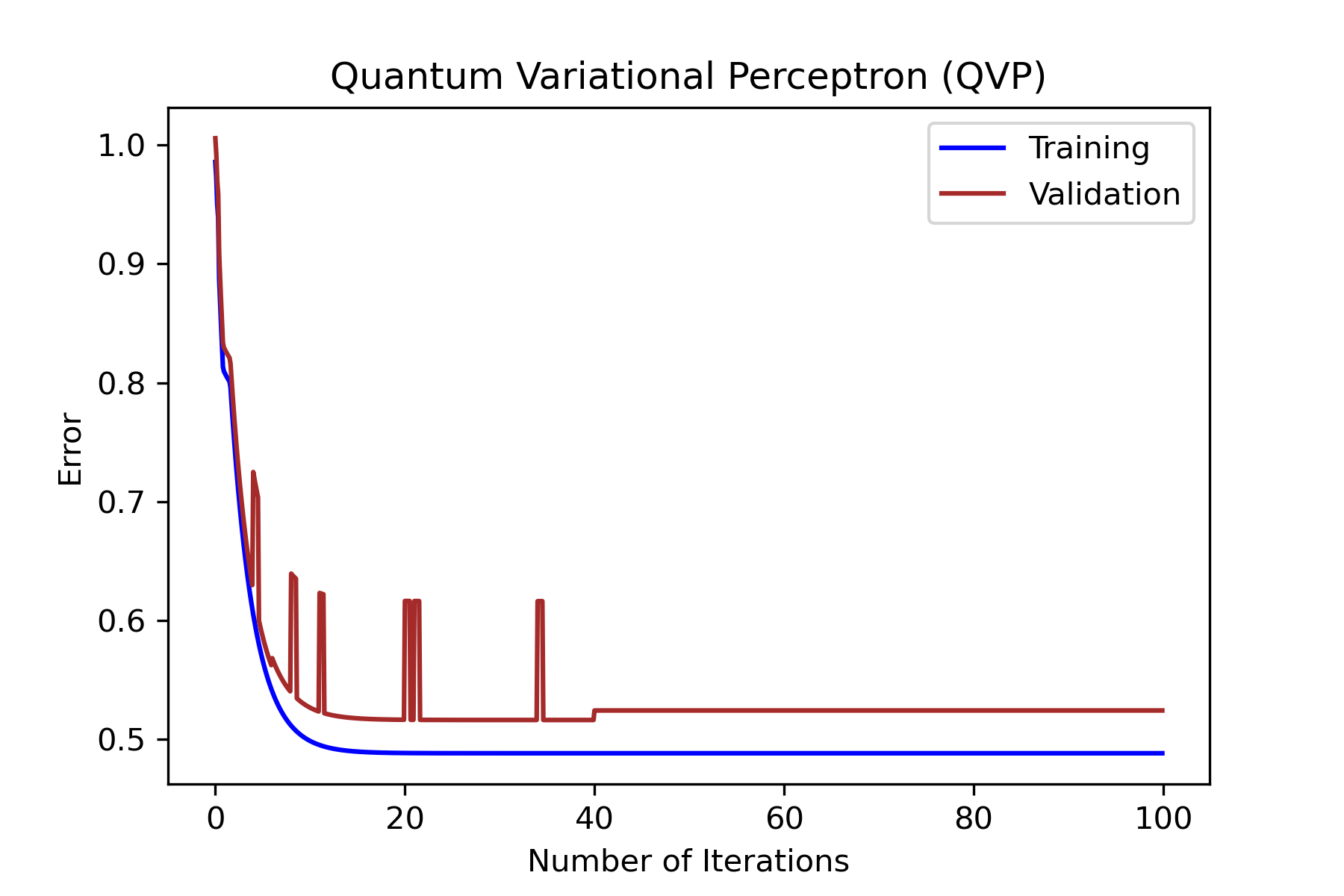}
  
\endminipage\hfill
\minipage{0.5\textwidth}%
  \includegraphics[width=\linewidth]{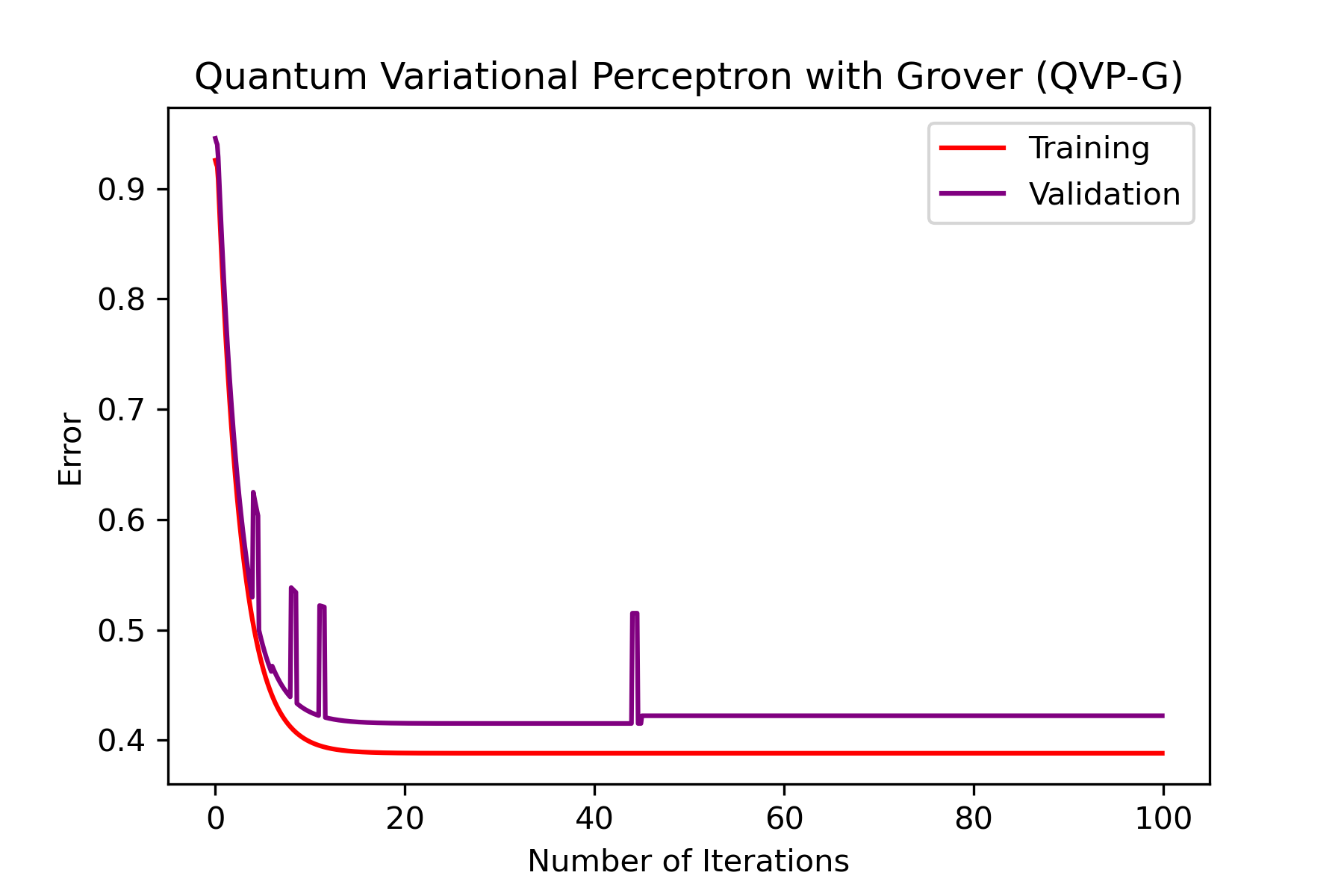}
  
\endminipage
\caption{Loss function of CP, GVP, and GVP with Grover}\label{loss}
\end{figure}

\indent Accordingly, we can assume that none of the three models is overfitting. The difference is that, generally, when a model reaches high validation accuracy in fewer iterations, it might be considered more efficient because it needs fewer data points to make accurate predictions. 
In this work, our quantum variational perceptron model and the quantum variational perceptron with Grover have a high validation accuracy of $99\%$ and reach that accuracy after a few iterations, indicating that they might be more efficient than the classical perceptron model.  
\newline
\indent The classical perceptron model has the lowest validation accuracy at $96\%$, but it reached this accuracy after most iterations, which shows that both quantum models are better. However, it may be more robust and generalizable to new data. 
\newline
\indent It is essential to mention that the number of iterations alone is insufficient to decide which model is better. We can explain this because of the use of quantum computers that, with quantum computing laws, have more powerful and efficient ways to predict the output. Furthermore, because of the use of the Grover circuit, our suggested model became in a configuration that made the measure of probability go smoother and faster, which led us to the expected value of our probability and weights in a speedy way to reach good accuracy. 
\newline
\indent Sometimes, it is crucial to stop the training when the model starts to overfit; this can be done by monitoring the loss or accuracy over time, which is why the next step is to compare the loss function.
\newline
\indent We use the loss function to optimize a quantum machine learning model or a classical one, measuring the difference between the model's predicted output and the actual output. We typically quantify this difference as an error, exactly the sum of errors made for each example in the data set; that is why we train the model to minimize this error.   
\newline
\indent The loss value implies how poorly or well a model behaves after each iteration of optimization, so we have tried to plot the loss function of the three methods: the classical perception, the quantum variational perceptron, and the quantum variational perceptron with Grover, figure \ref{loss} shows that the loss function of our approach converged well over the $100$ iterations over the other models (CP $\&$ QVP), and as known, a lower loss function value generally indicates better performance, as it means that the model's predictions are closer to the actual values. So the loss function of QVP-G ($\approx0.4$) would be considered the best performing, so we can assume that our method improves the model's performance.

\section{Conclusion}
\label{section:my4}
In this study, we proposed a novel model of a quantum perceptron that can achieve high accuracy to solve the classification task. First, we built our model using the variational circuit and Grover's algorithm. Then we demonstrated that our proposed model performs better by comparing its results to the variational quantum and classical perceptrons.
\newline
\indent Furthermore, our model's loss was minimal compared to the other models (CP $\&$ QVP). Those results proved that our model is accurate and perfectly fits some data. Hence, the use of Grover's algorithm in quantum machine learning, specifically the perceptron model, is promising since this approach can reduce the number of iterations needed to achieve the best accuracy. 
\newline
\indent In the future, we intend to develop other quantum perceptron models, evaluate their performance with the classification tasks, and use the proposed perceptron in real-world applications. Another exciting research path is to explore and implement other quantum models.

\end{document}